\documentclass{article}

\usepackage{amssymb}                    
\usepackage{amsmath}
\usepackage{graphicx}
\usepackage{textcomp}
\usepackage{natbib}
\usepackage{authblk}

\begin{document}
\title{Curvature, a mechanical link between the geometrical complexities of a fault.}
\author[1]{P. Romanet}
\author[2]{D. Sato}
\author[1]{R. Ando}
 \affil[1]{The University of Tokyo, 7-3-1 Hongo, Bunkyo-ku, Tokyo 113-8654, Japan}
  \affil[2]{Disaster Prevention Research Institute, Kyoto University, Gokasho, Uji, Kyoto 611-0011, Japan}

\maketitle

\section*{Abstract}
Many recent studies have tried to determine the influence of geometry of faults in earthquake mechanics. In this paper, we suggest a new interpretation of the effect of geometry on the stress on a fault. Starting from the representation theorem, which links the displacement in a medium to the slip distribution on its boundary, and assuming homogeneous infinite medium, a regularized boundary-element equation can be obtained. Using this equation, it is possible to separate the influence of geometry, as expressed by the curvatures and torsions of the field line of a dislocation on the fault surface, which multiply the slip, from the effect of the gradient of slip. This allows us to shed new light on the mechanical effects of geometrical complexities on the fault surface, with the key parameters being the curvatures and torsions of the slip field on the fault surface. We have used this new approach to explain further the false paradox between smooth-and-abrupt-bends (see \citet{sato2019}) as well as to re-interpret the effect of roughness on a fault.

\section{Introduction}
The geometry of a fault has an effect on the mechanics of earthquakes, as has been supported by many studies, including observations \citep{king1985,wesnousky2006,wesnousky2008,bletery2016}, theoretical work \citep{poliakov2002,rice2005}, and numerical modeling \citep{aochi2000a,bhat2004,oglesby2005,li2016,ando2017,romanet2018,ando2018}. Although it is now easy to study the consequences of geometry from numerical models, the underlying mechanics remain unsolved. All faults present some degree of geometrical complexities; it can be a curved subduction zone along the strike, like Sumatra or Cascadia, bends or kinks, or maybe even multiple faults and branches. Even at smaller scales, faults presents complexities. In fact it has been shown that faults are rough on scales covering more than nine orders of magnitude \citep{candela2012}. Knowing that geometrical complexities are present at all scales and that they do have effects, one natural question is whether  we can actually discriminate among all the geometrical complexities to see which one has the strongest effect. How exactly does the geometry come into play in the mechanics of an earthquake? In this paper we use elastodynamics to shed new light on these questions from an analytical perspective. 

Our starting point is to take advantage of a regularized boundary-element equation, first recalling some established results. We start from the representation theorem, which links the slip on a fault to the resulting displacement everywhere in the medium \citep{aki2002}:
\begin{equation}
\begin{split}
u_n(\mathbf{x},t) &=\int_{0}^{t}  \iint _{\Sigma} \Delta u_i(\mathbf{y},\tau)c_{ijpq} n_j (\mathbf{y})  \frac{\partial}{\partial y_q}  G_{np}(\mathbf{x},t-\tau;\mathbf{y},0) \mathrm{d}\Sigma(\mathbf{y}) \mathrm{d}\tau.
\label{representation_theorem}
\end{split}
\end{equation}
Here $u_n(\mathbf{x},t) $ represents the $n$-th component of the displacement, at location $\mathbf{x}$ and time $t$. $c_{ijpq}$ is the Hooke tensor, $n_j$ is the $j$-th component of the vector normal to the fault, $\Delta u_i(\mathbf{y},\tau)$ is the $i$-th component of the slip on the fault, and $G_{np}(\mathbf{x},t-\tau;\mathbf{y},0) $ is the $n$-th component of the Green's function, due to a delta function in component $p$ at location $\mathbf{y}$. Finally, $\Sigma$ represents the surface of the fault and $\tau$ the time history of the fault. From this equation, by assuming an elastic medium it is possible to derive the boundary-integral equation that relates the slip on the fault to the stresses everywhere in the medium. By an abuse of language, we will use the word "slip", to refer either to the shear slip or the opening in the following of the paper. 
Using Hooke's law, and noting that the stress $\sigma_{ab}$is given by:
\begin{equation}
\sigma_{ab} = c_{abcd} \left(      \frac{\partial u_d}{\partial x_c} + \frac{\partial u_c}{\partial x_d} \right),
\end{equation}

we can derive the boundary-integral equations after some algebra (\cite{tada1997,tada2000}):
\begin{equation}
\sigma_{ab}(\mathbf{x},t) =c_{abcd}\int_{0}^{t}d\tau  \iint _{\Sigma} \Delta u_i c_{ijpq} n_j \frac{\partial}{\partial y_q}\frac{\partial}{\partial x_d} G_{cp}d\Sigma.
\label{hypersingular}
\end{equation}
This equation is difficult to interpret because the slip is expressed in a general coordinate system, and also because, this equation is hypersingular \citep{martin1989} (this integral naturally diverges, but can be evaluated if we use Hadamard finite part definition of the integral). A way to be able to physically interpret this integral is to regularize it.

\section{Fundamental results}
 Following various authors (\cite{sladek1983},\cite{bonnet1999} p345), we introduce the tangential differential operator $D_{ij}=n_i\dfrac{\partial}{\partial x_j}-n_j\dfrac{\partial}{\partial x_i}$, where $n_i$ represents the i-th componant of the  vector $\mathbf{n}$ normal to the surface. Using this operator and assuming a linear, elastic, infinite, homogenous medium allows to perform the regularization of eq. \eqref{hypersingular} by using integration by parts (\citet{sladek1984,bonnet1999} p 175). More details are given in appendix A. For the connection with the local coordinate system that has frequently been adopted in the rock mechanics literature (e.g., \citet{tada1997,tada2000}) please refer to \citet{sato2019}.
\begin{equation}
\begin{split}
\sigma_{ab}(\mathbf{x},t) &=c_{abcd}\int_{0}^{t}d\tau \iint _{\Sigma}c_{ijpq}D_{jd} [\Delta u_i] G_{cp}   d\Sigma 
 \\ &-c_{abcd}\int_{0}^{t}d\tau  \iint _{\Sigma} \rho\frac{\partial^2}{\partial t^2}\Delta u_in_d  G_{ic} d\Sigma 
\end{split}
\label{reg1}
\end{equation}

\begin{figure}
\centering
\includegraphics[width=90mm]{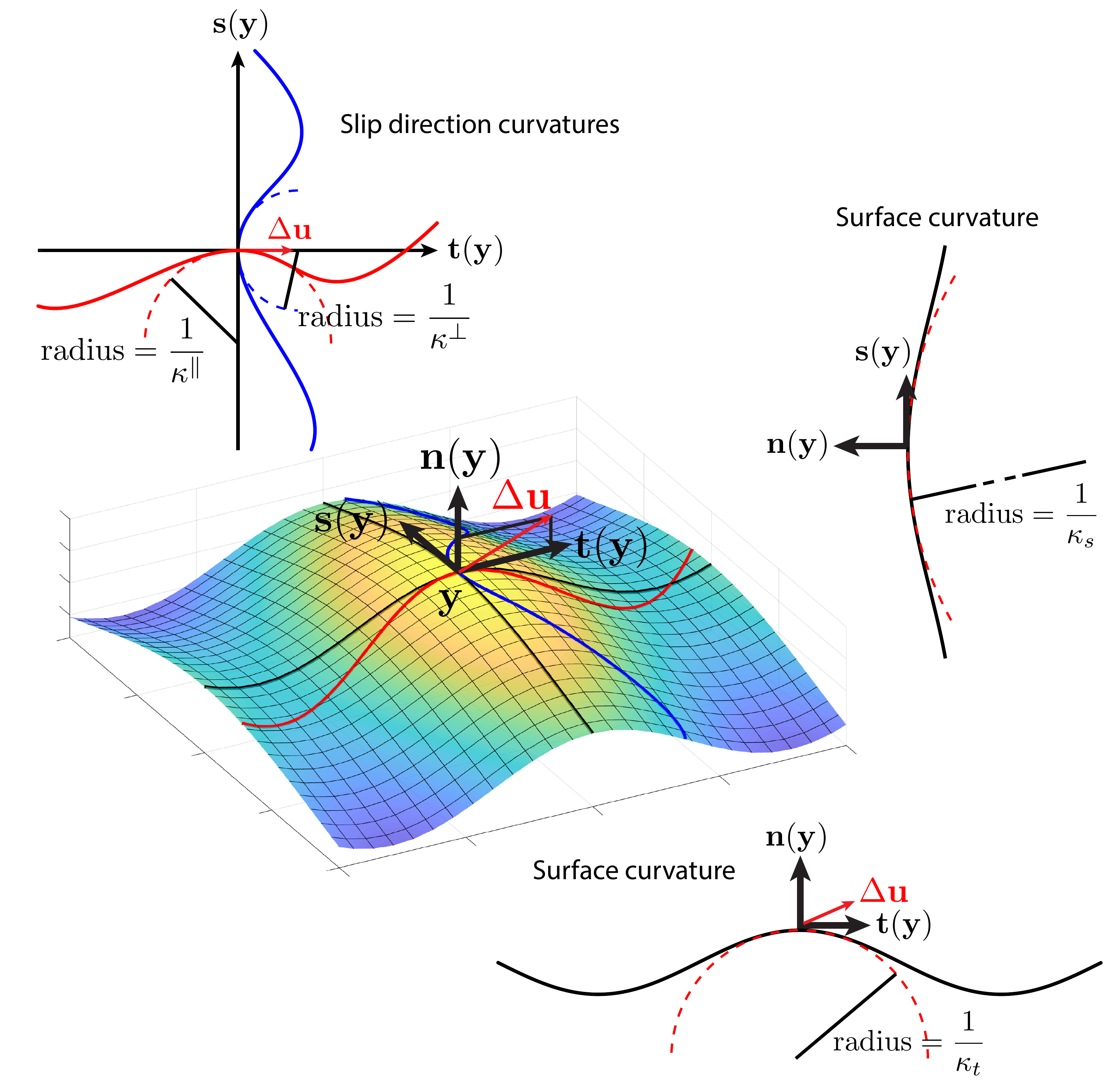}
\caption{Figure showing the four curvatures of a fault in 3D medium. The curve in red is the field line of tangential slip, as projected onto the fault. The blue line is the field line perpendicular to the field line of tangential slip at every point. The color of the surface represents the elevation.}
\label{surface}
\end{figure}

This equation can be interpreted by projecting the derivatives and the slip onto a local coordinate system that is chosen as follows (see Fig. 1): the tangential vector $\mathbf{t}$ is parallel to the projection of the slip vector $\Delta \mathbf{u}$ on the surface at any point, and $\mathbf{s}$ is the perpendicular cotangent vector; the slip vector has no component on $\mathbf{s}$.
The last step is to separate the contributions from shear slip and opening: $\Delta u_i=t_i \Delta u^t + n_i \Delta u ^n$.
By doing so, it is possible to write the integral in eq. \eqref{reg1} in a form involving different contributions with physical meanings: the effects of the slip gradients (the second and third lines of eq. \ref{main_eq} below), the effects of the curvatures ($\kappa^t$, $\kappa^s$, $\kappa^{\parallel}$, and $\kappa^{\bot}$ ) and the torsions ($\tau^t$ and $\tau^s$) of the slip field lines on the fault surface multiplying the slip (the fourth and fifth lines of eq. \ref{main_eq}), and the effect of inertia (the sixth line of eq. \ref{main_eq}).
\clearpage
\begin{equation}
\begin{split}
&\sigma_{ab}(\mathbf{x},t) = \\
&\text{\underline{Gradient of slip:}}\\
&c_{abcd}\int_{0}^{t}  \iint _{\Sigma}c_{ijpq}\frac{\partial}{\partial x_q} G_{cp} (n_d t_j- n_j t_d) \\ 
&\qquad \qquad  \qquad \qquad \qquad \left[  t_i \frac{\partial}{\partial y^t} \Delta u^t+n_i \frac{\partial}{\partial y^t}\Delta u^n\right]   d\Sigma d\tau \\
+&c_{abcd}\int_{0}^{t}  \iint _{\Sigma}c_{ijpq}\frac{\partial}{\partial x_q} G_{cp}(n_d s_j- n_j s_d) \\
& \qquad   \qquad \qquad \qquad \qquad  \left[ t_i \frac{\partial}{\partial y^s} \Delta u^t+n_i \frac{\partial}{\partial y^s}\Delta u^n   \right]   d\Sigma  d\tau\\
&\text{\underline{Curvature and torsion times slip:}}\\
+&c_{abcd}\int_{0}^{t}  \iint _{\Sigma}c_{ijpq}\frac{\partial}{\partial x_q} G_{cp}(n_d t_j- n_j t_d) \\
& \qquad \qquad  \left[  (\kappa^t  n_i+\kappa^{\parallel} s_i )\Delta u^t +  (\tau^t s_i - \kappa^t t_i )\Delta u^n  \right]   d\Sigma d\tau  \\
+&c_{abcd}\int_{0}^{t}  \iint _{\Sigma}c_{ijpq}\frac{\partial}{\partial x_q} G_{cp} (n_d s_j- n_j s_d)  \\
&\qquad  \qquad  \left[ (\tau^s n_i + \kappa^{\bot} s_i)  \Delta u^t  - (\kappa^s s_i +\tau^s t_i )\Delta u^n \right]   d\Sigma d\tau \\
&\text{\underline{Inertia term:}}\\
-&c_{abcd}\int_{0}^{t}\iint _{\Sigma} \frac{\partial^2}{\partial t^2}\Delta u_i(\mathbf{y},\tau)n_d \rho G_{ic} d\Sigma  d\tau
\label{main_eq}
\end{split}
\end{equation}

This equation allows us to understand better the different sources of stress due to a slip distribution on a fault. The last term is the inertial term, which is due to the second derivative of slip with time; it disappears when the fault is no longer slipping for a sufficiently long time. The other terms are each constructed in a similar manner:
\begin{equation}
\begin{split}
&\int_{0}^{t}  \iint _{\Sigma}\underbrace{c_{abcd}c_{ijpq}\frac{\partial}{\partial x_q} G_{cp}(n_d t_j- n_j t_d)}_{\text{Kernel}} \\ 
&\qquad \qquad\underbrace{\left[  (\kappa^t  n_i+\kappa^{\parallel} s_i )\Delta u^t +  (\tau^t s_i - \kappa^t t_i )\Delta u^n  \right] \vphantom{c_{abcd}c_{ijpq}\frac{\partial}{\partial x_q} G_{cp}(n_d t_j- n_j t_d)}}_{\text{Source}} d\Sigma d\tau .
\end{split}
\end{equation}

There are two different kernels, which differ only in the projection onto the fault, and two different sources. \textbf{One novelty of the present work is that the source is not only the gradient of the slip along the fault, called the dislocation, but in addition another kind of source arises from the curvatures and torsion that multiply the slip.} This is not the first time that a curvature term has been found in the development of a boundary-element equation. \cite{koller1992} found a similar term for the dynamic antiplane case in which the boundary-element equation is fully regularized (i.e., there is no Cauchy integral to evaluate). However, our work is truly different from the curvature term as treated in the anti-plane case of \citet{koller1992}, where the traction becomes null when the dislocation goes to zero. The curvature and torsion terms in our work do not vanish when the dislocation is null. These curvatures and torsions are fully defined by the field line of the slip on the fault surface (e.g., the red line in Fig. \ref{surface}). Interestingly, the kernels for both source contributions are exactly the same. The interpretation of the curvatures and torsions is as follows (see Fig. \ref{surface}): two are related to the curvature of the fault surface, respectively parallel to the direction of slip projected onto the fault ($\kappa^t$), and perpendicular to that projection of the slip direction ($\kappa^s$). We call these two terms "fault-surface curvatures", because they are related to the geometry of the fault. The other two curvatures are related to changes in the direction of slip projected onto the fault: the change in the direction of the slip itself ($\kappa^{\parallel}$) and the change in the perpendicular direction ($\kappa^{\bot}$); hence, we call them "slip-direction curvatures" (strictly speaking, we should call them geodetic curvatures). The two last terms are the torsions $\tau^t$ and $\tau^s$ (strictly speaking, the relative torsions). They can be thought of as the conjugate effect of the geometry of the fault and the change in direction of the slip.  The stress due to tangential slip depends only on one fault-surface curvature, the one that is in the direction of the slip. It also depends on the two slip-direction curvatures and one torsion. On the other hand, the stress due to opening does not depend on the slip-direction curvatures, but only on the fault-surface curvatures and torsions. This provides a fundamental difference between the non-planar geometry of opening and shear cracks: although other inertial and gradient contributions can be exactly the same, the contribution from the curvatures and torsions is fundamentally different.

\section{Results and discussion}

\begin{figure}
\centering
\includegraphics[width=80mm]{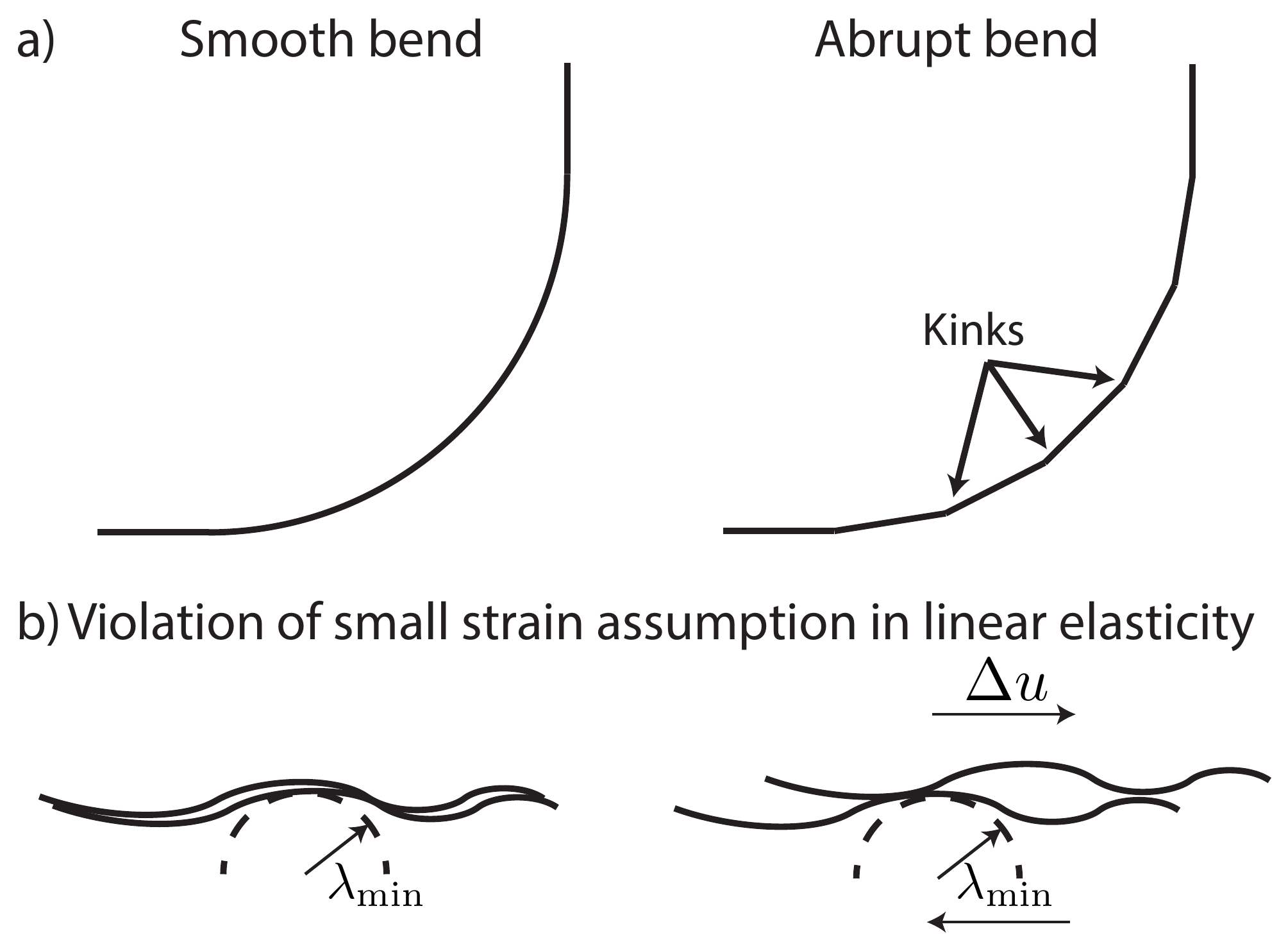}
\caption{a) Illustration of the smooth-and-abrupt-bend paradox. Calculation of the stresses yields different results for a smooth bend (left panel) or an abrupt bend (right panel), even in the limit of infinitesimally small segments. b) Sketch showing the need for a minimum lengthscale when considering linear elasticity, which is required by the small-strain assumption; the ratio $\Delta u / \lambda_{\text{min}}$ must remain small.}
\label{discussion}
\end{figure}

\subsection{Relationship to previous work}
The work of \cite{tada1997,tada2000} was done independently of the regularization process using the tangential differential operator. They regularized the boundary-integral equations by projecting the derivative into the local coordinate system and then performing an integration by parts. However, they made a mistake in performing the integration by parts, because they neglected the derivative of the local coordinate system. Their work was corrected by \citet{sato2019}. The apparent smooth-and-abrupt-bend paradox \citep{tada1996} that results from this omission states that for two bends, one connected with a smooth curve and another connected with straight-line segments (see. Fig. \ref{discussion}.a), the stress calculation yields different results, even in the limit of infinitesimally small straight segments. Through numerical calculation, \citet{sato2019} showed that this apparent paradox is not correct.  This also means that there is no problem in using flat elements to model a curve fault. The work done by \cite{aochi2000b}, in which they first calculate the effect of a constant slip rate on a small flat element and then assemble these flat elements (by rotation and translation) to create a complex geometry therefore encounters no problem regarding the smooth-and-abrupt-bend paradox. The use of the tangential differential operator avoid the choice of a local basis along the fault, so this provides an unified view of the derivation, independently of the choice of the coordinate system. 

\subsection{A geometrical effect}
The effects of the curvatures and torsions of the fault surface are truly geometrical effects. This is quite different from the effect of the slip gradient, which can easily be positive or negative on the fault surface and which can evolve with time and eventually reverse sign to fulfill the requirement due to friction applying in the fault. However, if we assume that the geometry of the fault and the direction of the shear slip do not change, the curvature term that multiplies the  slip can only evolve monotonically. This means that we expect high stress concentrations at all geometrical complexities, where "geometrical complexities" stands for any part of the fault where the local curvature is high. In addition to damage, breaks in the free surface, another means for decreasing this stress is actually to decrease the fault-surface curvatures, i.e., to make the fault become flatter. Qualitatively, this effect is observed for mature faults with high slip; they tend to be flatter than the youngest faults. 

 If one is interested only in the displacement produced by a slip distribution on a fault, there is no such effect of curvature (see eq. \eqref{representation_theorem}). A way to understand this, is to think of the curvature terms multiplied by the slip as a strain (the same way the gradient of slip is also a strain). The geometry still has an influence through the Green's function, but the curvature is not a source term. Because of this, the assumption of local planarity of a fault for slip inversion is probably a good first approximation. However, as soon as one wants to calculate the strains or stresses, the planar-fault assumption and the true geometrically complex fault gives rise to completely different results: the curvature contribution is neglected, although we have shown that this term can be dominant. This means that one must take care in calculating the stress from slip inversion. Locally on the fault, the true stress state (partly because of the complex local geometry) is probably very different from the one calculated by assuming a locally planar fault. In other words, it is not because two slip distributions are similar that the strain calculated from them is close, because it also depends strongly on the geometry. Another direct consequence is that dynamic (as opposed to kinematic) inversions of the slip on faults are probably very difficult to implement other than statistically, as there is no way to know the exact geometry of a fault.

\subsection{In plane elasticity with no opening}
The previous result is very general, and it holds for any fault geometry, including multiple faults, in the dynamic case, and it allows for shear slip as well as for opening. To explore this new finding, let us simplify this expression to in-plane elastostatic problem with no opening. In 2D, the torsions and slip direction curvatures disappear, and only one surface curvature remains. In this particular case, the stress in the medium depends only on one curvature mutiplying the shear slip and on the gradient of shear slip along the fault:
\begin{equation}
\begin{split}
&\sigma_{ij}(\mathbf{x}) = \\
&\int_C K_{ij}^{\text{curv}}(\mathbf{x}-\mathbf{y}) \kappa^t\Delta u^t d L(\mathbf{y})+\int_C K_{ij}^{\text{grad}}(\mathbf{x}-\mathbf{y}) \frac{\partial}{\partial y^t}\Delta u^t d L(\mathbf{y}).
\label{stress2D}
\end{split}
\end{equation}
The two kernels $ K_{ij}^{\text{curv}}(\mathbf{x}-\mathbf{y})$ and $K_{ij}^{\text{grad}}(\mathbf{x}-\mathbf{y}) $ are given in appendix \ref{stressB1}. The difference from the paper of \cite{tada1997}, is the term involving the curvature in eq. \eqref{stress2D}, because that reference neglected the derivative of the local coordinate system.

\subsubsection{Existence of a lengthscale related to geometry in numerical models (small strain elasticity)}
If we want the stress to remain finite in eq. \eqref{stress2D}, for any geometry, we need to assume a finite curvature along the fault. Otherwise, the slip must vanish at the point of divergent curvature. This also means that, if we want the stress to remain finite at a kink, the only solution is that the slip must be null at the position of the kink. This limitation arises from the fact that these equations were derived in the small-strain approximation (i.e.: linear elasticity): two asperities in contact cannot slide past other each other, as this would break the assumption of small-strains (see Fig. \ref{discussion}.b). In modeling, we are limited by the fact we want to have some slip on the fault, and this necessarily introduces a minimum lengthscale $\lambda_{\text{min}}$ into the system. In practice, the slip has to be much smaller than the curvature: $\Delta u\ll \lambda_{\text{min}} $. Obviously, this is a rather arbitrary limit  to the small asperities in modeling, so how can we deal with this lengthscale? One way would be to go beyond the small-strain approximation and use non-linear elasticity. But then the equation become much more complex to solve numerically. Another way would be to incorporate some of the small-scale asperities into the friction law. Such small-lengthscale asperities probably have very localized effects, and therefore this may be sufficient for numerical modeling without using the more complicated tools of non-linear elasticity. Future research is needed to understand the effects of this minimum lengthscale.  
Another related topic is that one must take care not to break the small-strain assumption of linear elasticity when incorporating roughness in modeling. If we set the minimum roughness to be $\lambda_{\text{min}}=10^{-2}\mathrm{mm}$, and if we assume that linear elasticity holds for maximum strains of $5\%$, this means that the displacement on a fault  can be no more than $5\times 10^{-4}\mathrm{mm}$.

\subsubsection{Bends and kinks}
\begin{figure*}
\centering
\includegraphics[width=\textwidth]{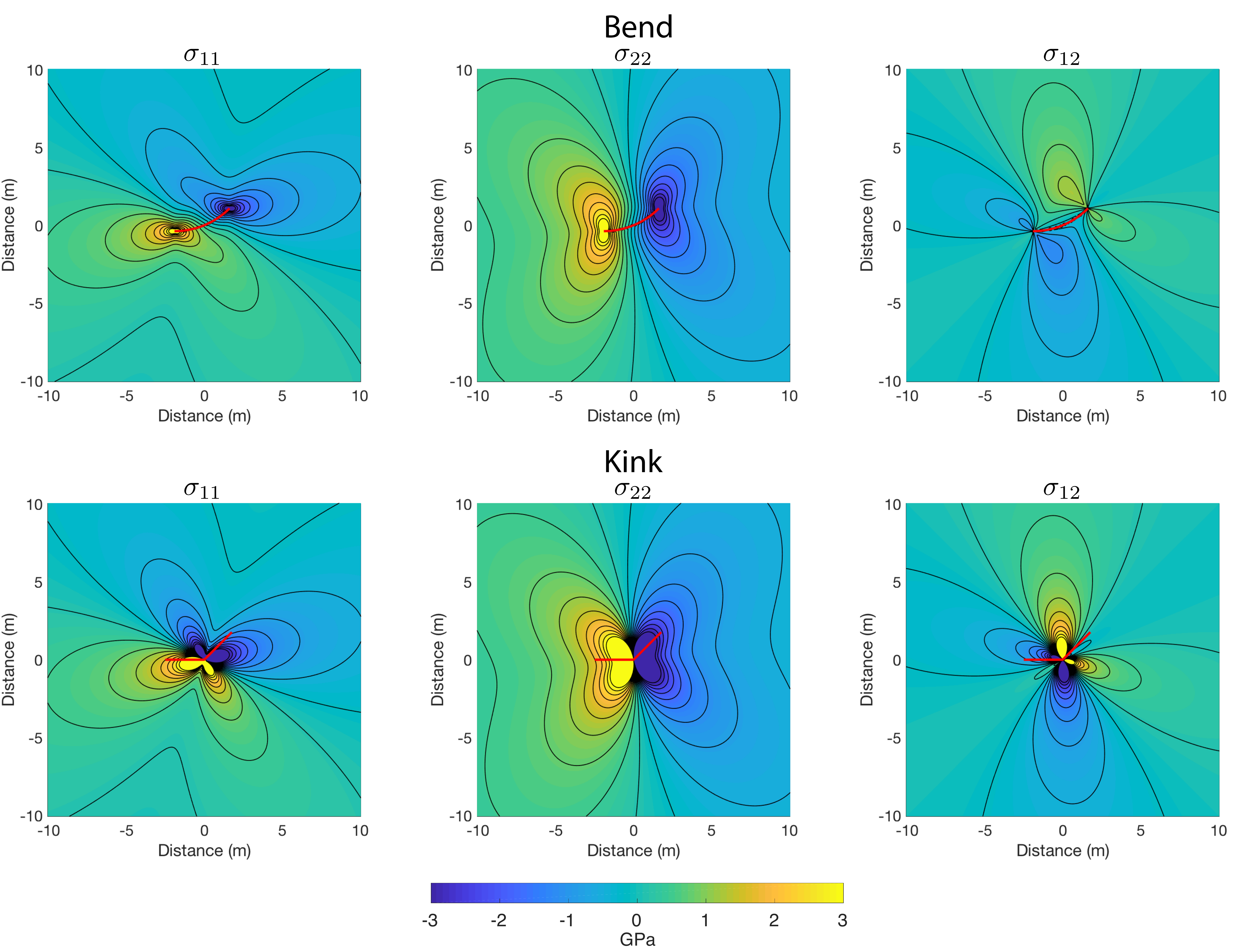}
\caption{Stress in a medium, due to a  bend (upper panels), and due to an abrupt bend (lower panels). The slip is constant over the fault. We removed the effect of gradient at the edges of the fault. The angle of aperture is 45\textdegree. }
\label{stress}
\end{figure*}

\begin{figure*}
\centering
\includegraphics[width=\textwidth]{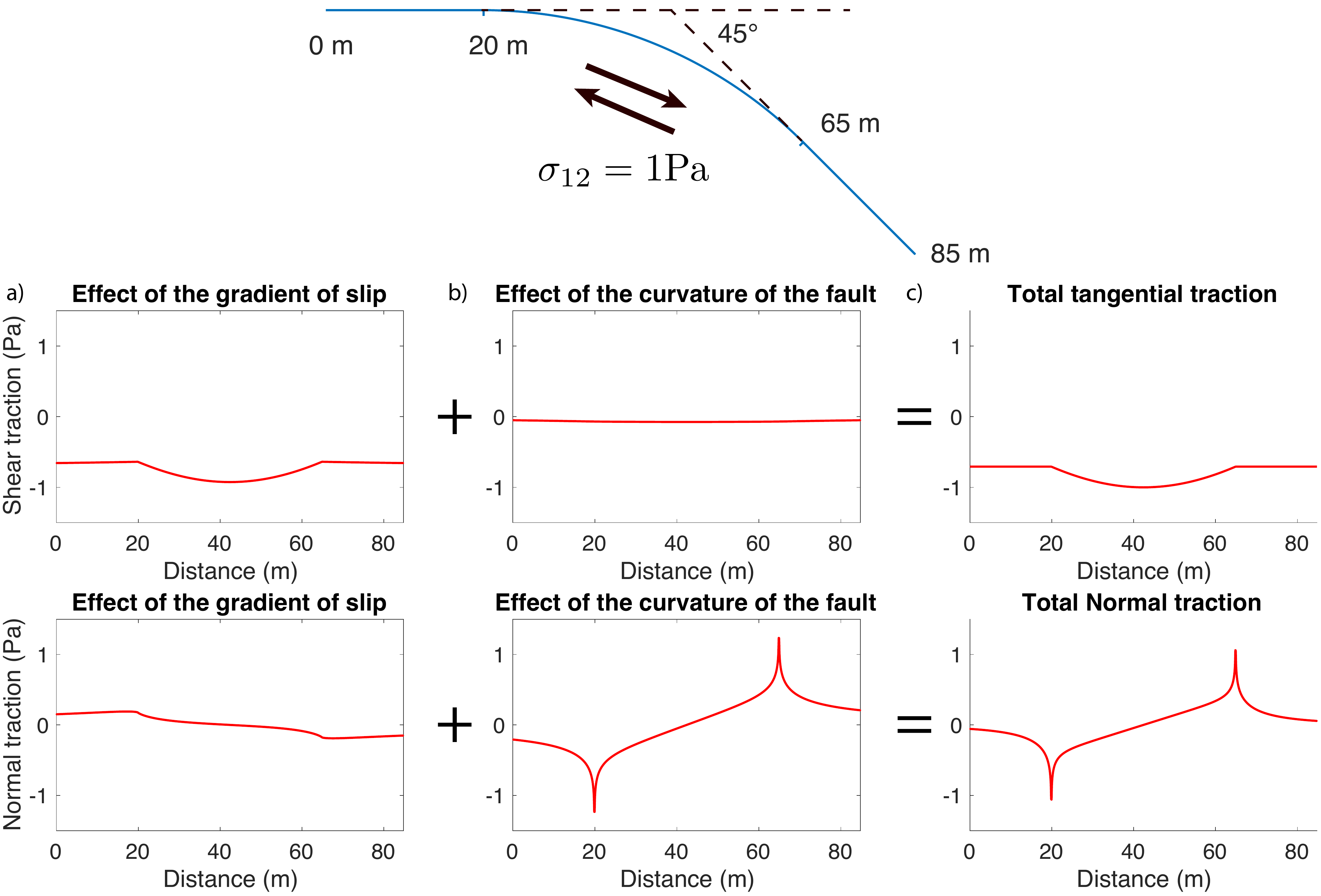}
\caption{The effect of the different contributions of source terms on a curved fault (upper panel), with the assumption that the shear traction is null on the fault. a) Effect of the gradient of slip along the fault on the normal and shear traction. b). Effect of the curvature that multiplies slip along  the fault on the normal and shear traction.. c) Total effects on the normal and shear traction on the fault.}
\label{normal_traction}
\end{figure*}

The difference between bends and kinks is an interesting question in the community of fault modeling. Our work allows us to make the link between the two. As stated earlier, the only natural solution when a kink arises on a single fault is for the slip to be zero at the kink. In Fig. \ref{stress}, we compare the effects of a kink and a smooth bend on the stress. For this purpose, we calculated the effect of a constant tangential slip on a kink and on a bend (the effect of the gradient of slip along the fault is therefore null). This figure shows that the contribution of a kink and a bend on the far field is the same, which provide further evidence to show that the smooth-and-abrupt-bend paradox is wrong \citep{sato2019}. Imagine that we zoom out on the bend, so that we cannot distinguish it anymore; then the solution will be similar to that for a kink. This is rather good news, as it respects the Saint Venant's principle that "the difference between the effects of two different but statically equivalent loads becomes very small at sufficiently large distances from load". If we look at the near field, however, the kink has a singular stress, which the smooth bend does not have. This arises because we imposed a non-zero slip at the kink, which breaks the small-strain assumption of linear elasticity. From the distinction between the effect of the gradient of slip and the effect of the curvature that multiplies the slip, we also can analyze again Figure 2 of  \cite{tada1996} which shows the normal traction along a bended crack (the total shear stress on the fault, including the elastic and loading shear stress, has to be zero). This figure is similar to Fig. \ref{normal_traction}, where we distinguish between the traction arising from the gradient of slip and the traction arising from the curvature. The difference between these results arises from the fact that they neglected the curvature term. Note that in this example, the normal traction comes mainly from the curvature term, and the gradient of slip has a smaller influence. Conversely, the effect of curvature on the shear stress is small, and it opposes movement. In this case, the effect of the gradient of slip is preponderant. In summary, a kink can be seen as the limiting case of an infinitesimally small smooth bend. And the terms that matter for understanding the difference between an abrupt bend and a smooth bend is really the curvature term.


\subsubsection{Rough faults}
\begin{figure*}
\centering
\includegraphics[width=\textwidth]{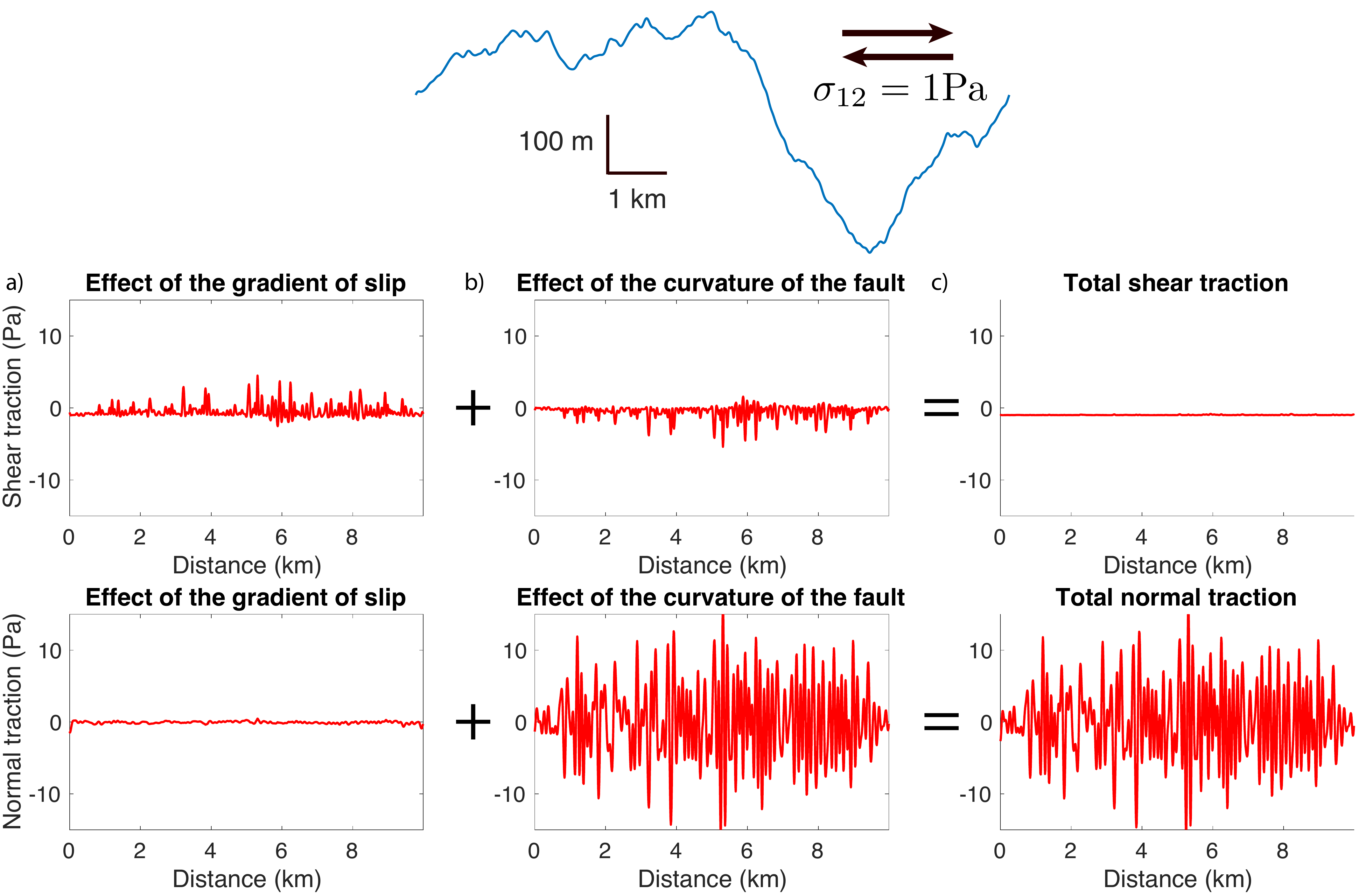}
\caption{The effect of the different contributions of source terms on a rough fault (upper panel), with the assumption that the shear traction is null on the fault. a) Effect of the gradient of slip along the fault on the normal and shear traction. b). Effect of the curvature that multiplies slip along  the fault on the normal and shear traction.. c) Total effects on the normal and shear traction on the fault.}
\label{roughCrack}
\end{figure*}

Rough faults have received heightened attention recently because of their effects in earthquake dynamics. It is known that faults actually present roughness at all scales (\cite{power1987, power1991, renard2006, candela2009,candela2012}). Analytical studies of roughness by boundary perturbations always find a scaling of stress that is proportional to slip, the inverse of the minimum lengthscale of roughness, and the shear modulus (\cite{dunham2011b,fang2013}). This is in contradiction with the boundary method developed in \cite{tada1997,tada2000}. Our results actually confirms this scaling in a more general way, as we found that it is possible to separate the two contributions. To prevent the equation \eqref{stress2D}  from diverging, it now appears clearly that a minimum roughness must be set.This also provides an explanation for the difference in roughness measurements in the direction of slip, and perpendicular to the direction of slip \citep{power1987,power1991,lee1996}. Indeed, we can assume that faults rarely open, so that there is only shear slip on it. The resulting stresses from the shear slip depend only on the curvature in the same direction as the slip ($\kappa^t$). The stress increasing with ongoing shear slip, it makes sense that the roughness, by some damage process, will vanish with ongoing slip (\cite{sagy2007,brodsky2011}). Figure \ref{roughCrack} shows the contribution of the gradient of slip and the curvature, for a rough crack with an amplitude-to-wavelength ratio $\alpha =10^{-2}$, that is loaded with pure shear. Similar to the bend in Fig. \ref{normal_traction}, the shear traction depends mainly on the gradient of slip, while the normal traction depends mainly on the curvature that multiplies the slip. The normal stress shows large variations, more than 10 times the initial shear stress drop. The curvature term also seems to correspond to the drag stress found by \cite{fang2013}: the effect of curvature on shear stress opposes the movement of the fault.  In this case, the effect of geometry cannot be neglected. As discussed previously, a purely rough fault without a minimum lengthscale $\lambda_{\text{min}}$ can not be modeled using linear elasticity. The curvature would be infinite for such a fault, and therefore the only solution would be that the slip is null.

\section{Conclusion}
In this paper we have demonstrated the key role of fault curvature in earthquake mechanics. By using regularized boundary-integral equations, we are able to separate the stress contribution into several terms: terms that are proportional to the slip gradient and terms that are proportional both to slip and to the curvatures of the fault. This helps us to understand better the mechanics of fault systems. We have also provided additional arguments which demonstrates that there is no such a thing as a paradox of smooth-and-abrupt-bends (\cite{tada1996, aki2002}, p591, \citet{sato2019}). We have also used this results to generalize recent finding about roughness stress drag and its particular scaling, proportional to slip and the inverse of a minimum lengthscale. We find that the effect of curvature, and therefore geometry is not negligible, even for small departures from a planar fault. The potential applications of this finding are large, and for the first time this work has provided a quantitative way of measuring the effect of geometry of faults on the stresses. This finding can help to understand better, for example, the effect of assuming a planar fault compared to a non-planar fault. This can help to argue that bends and kinks are acting as a barrier, for example, in the case of paleoseismology. This can also help to understand the variation of roughness along the faults, depending on the direction of slip. Finally, we find that the curvature of a fault is the key physical parameter that unifies and defines a complex geometry.

\section*{Author contribution statement}
P.R. wrote the manuscript. D.S. initially found the regularization technics of the boundary-integral equation. P.R. found the effect of curvature and the implications in 2D. Both D.S. and P.R. participated in combining the two works and discussed the results. R.A. initiated the project and found the mistake in the work of \citet{tada1997}. Finally, all the three authors have read and approved the present manuscript.

\section*{Ackowledgement}
P.R would like to acknowledge valuable discussions with Dr. Bhat and Dr. Aochi. P.R. thanks greatly Dr. Ide for allowing him to come to the University of Tokyo. We greatly thank Dr. Bonnet for drawing us attention on the use of the tangential differential operator. P.R also thanks greatly Prof. Madariaga for comments it gets while he was writing this article. The matlab codes used to generate the figures are in supplementary material. This work was partially supported by JSPS KAKENHI 18KK0095 and 19K04031. P.R. received support from KAKENHI 16H02219. D.S was partially supported by JSPS/MEXT KAKENHI Grant Numbers JP26109007.

\appendix

\section{Derivation of regulatised boundary-integral equations} \label{sec.derivation}
\subsection{Green's function}
Here, we recall some properties of the Green's function of homogeneous linear elastic medium. The Green's function have some symmetry properties:
\begin{equation}
G_{ij}(\mathbf{x},t,\mathbf{y},\tau)  = G_{ij}(\mathbf{x}-\mathbf{y},t-\tau,0,0) 
\end{equation}
That leads to the following property for the derivative of the Green function:
\begin{equation}
\frac{\partial}{\partial x_q}G_{ij}(\mathbf{x},t,\mathbf{y},\tau) = -\frac{\partial}{\partial y_q}G_{ij}(\mathbf{x},t,\mathbf{y},\tau) 
\label{inversionDeriv}
\end{equation}
The Green's function also satisfied the momemtum balance equation:
\begin{equation}
\begin{split}
\rho \frac{\partial^2}{\partial t^2}G_{in} &= c_{ijpq}  \frac{\partial}{\partial x_j} \frac{\partial}{\partial x_q} G_{pn} 
\end{split}
\end{equation}

\subsection{Regularisation of the representation theorem}
In this section, we present a regularisation of the representation theorem for general fault geometry following \citet{bonnet1999}, p174-176. For that purpose, the tangential differential operator $D_{ij}[\sim]$ will be introduced \citep{sladek1983,bonnet1999}:
\begin{equation}
D_{ij}[f] = n_i \frac{\partial}{\partial x_j}f- n_j \frac{\partial}{\partial x_i}f
\end{equation}
Where $f$ is the function at which the tangential differential operator is applied.

Let us start from the representation theorem:
\begin{equation}
u_n(\mathbf{x},t) =\int_{0}^{t}d\tau  \iint _{\Sigma} \Delta u_i(\mathbf{y},\tau)c_{ijpq} n_j \frac{\partial}{\partial y_q} G_{np}(\mathbf{x},t-\tau;\mathbf{y},0) d\Sigma
\end{equation}
Using the strain stress relationship,
\begin{equation}
\sigma_{ab} = c_{abcd} \frac{1}{2}\left(      \frac{\partial u_d}{\partial x_c} + \frac{\partial u_c}{\partial x_d} \right)
\end{equation}
It is possible to derive the stress: 
\begin{equation}
\begin{split}
\sigma_{ab}(\mathbf{x},t) &=c_{abcd}\int_{0}^{t}d\tau \iint _{\Sigma} \Delta u_i(\mathbf{y},\tau) \\
& \ c_{ijpq} n_j \frac{\partial}{\partial y_q}\frac{\partial}{\partial x_d} G_{cp}(\mathbf{x},t-\tau;\mathbf{y},0) d\Sigma
\end{split}
\end{equation}
If now we use the property of Green's function \eqref{inversionDeriv}, the tangential differential operator and the equation of motion, we can show that:
\begin{equation}
\begin{split}
c_{ijpq} n_j \frac{\partial}{\partial x_d} \frac{\partial}{\partial y_q} G_{cp}  &=  -c_{ijpq} n_j \frac{\partial}{\partial x_d} \frac{\partial}{\partial x_q} G_{cp} \\ 
 &=- c_{ijpq} D_{jd}  \frac{\partial}{\partial x_q} G_{cp} - c_{ijpq} n_d \frac{\partial}{\partial x_j} \frac{\partial}{\partial x_q} G_{cp}  \\
 &= - c_{ijpq} D_{jd}  \frac{\partial}{\partial x_q} G_{cp} -  n_d \rho \frac{\partial^2}{\partial t^2} G_{ic}   
\end{split}
\end{equation}

We can rewrite the stress as following:

\begin{equation}
\begin{split}
\sigma_{ab}(\mathbf{x},t) &=c_{abcd}\int_{0}^{t}d\tau  \iint _{\Sigma} \Delta u_i(\mathbf{y},\tau)c_{ijpq}D_{jd}[G_{cp} ] d\Sigma  \\ 
&-c_{abcd}\int_{0}^{t}d\tau \iint _{\Sigma} -\Delta u_i(\mathbf{y},\tau)n_d \rho \frac{\partial^2}{\partial t^2} G_{ic} d\Sigma
\end{split}
\end{equation}
After performing integration by parts it comes:
\begin{equation}
\begin{split}
\sigma_{ab}(\mathbf{x},t) &=c_{abcd}\int_{0}^{t}d\tau\iint _{\Sigma}c_{ijpq}D_{jd} [\Delta u_i(\mathbf{y},\tau)] G_{cp}   d\Sigma 
 \\ &-c_{abcd}\int_{0}^{t}d\tau \iint _{\Sigma} \frac{\partial^2}{\partial t^2}\Delta u_i(\mathbf{y},\tau)n_d \rho G_{ic} d\Sigma 
\end{split}
\label{dev0}
\end{equation}

Where we use the fact that:
\begin{equation}
\begin{split}
\int_{0}^{t}d\tau  \iint _{\Sigma}D_{jd} [c_{abcd}c_{ijpq}\Delta u_i(\mathbf{y},\tau)G_{cp} ]  d\Sigma =0
\end{split}
\end{equation}
This is derived from Stoke's theorem (\cite{bonnet1999},p346). This regularised boundary-equation (eq. \eqref{dev0}) matches with the expression of \citet{sato2019}, where instead of using the tangential differential operator, the projection of the derivatives on a local basis on the fault is used.

\subsection{Derivation of curvature and torsion terms}
Using Darboux frame, it can be shown that:
\begin{equation}
\begin{split}
\frac{\partial}{\partial y^t} \mathbf{t}(\mathbf{y}) &= \kappa^t \mathbf{n}(\mathbf{y}) + \kappa^{\parallel} \mathbf{s}(\mathbf{y}) \\
\frac{\partial}{\partial y^t} \mathbf{n}(\mathbf{y}) &= -\kappa^t \mathbf{t}(\mathbf{y})+\tau^t \mathbf{s}(\mathbf{y}) \\
\frac{\partial}{\partial y^s} \mathbf{t}(\mathbf{y}) &= \tau^s \mathbf{n}(\mathbf{y}) + \kappa^{\bot} \mathbf{s}(\mathbf{y}) \\
\frac{\partial}{\partial y^s} \mathbf{n}(\mathbf{y}) &=-\tau^s \mathbf{t}(\mathbf{y}) - \kappa^s \mathbf{s}(\mathbf{y})
\end{split}
\end{equation}

\subsection{Separating the two contributions of slip: normal and tangential slip}
If we write the slip in term of shear slip and opening:
\begin{equation}
\Delta u_i = t_i \Delta u^t + n_i \Delta u^n
\end{equation}
It is possible to write the derivatives of the slip as: 
\begin{equation}
\begin{split}
\frac{\partial}{\partial y^t}\Delta u_i  &=\Delta u^t  \frac{\partial}{\partial y^t}t_i + t_i \frac{\partial}{\partial y^t} \Delta u^t \\
&+ \Delta u^n\frac{\partial}{\partial y^t}n_i  +  n_i \frac{\partial}{\partial y^t} \Delta u^n \\
&=(\kappa^t  n_i+\kappa^{\parallel} s_i )\Delta u^t  + t_i \frac{\partial}{\partial y^t} \Delta u^t \\
&+ (\tau^t s_i - \kappa^t t_i )\Delta u^n  +  n_i \frac{\partial}{\partial y^t} \Delta u^n
\end{split}
\label{dev1}
\end{equation}
And 
\begin{equation}
\begin{split}
\frac{\partial}{\partial y^s}\Delta u_i  &=\Delta u^t  \frac{\partial}{\partial y^s}t_i + t_i \frac{\partial}{\partial y^s} \Delta u^t \\
&+ \Delta u^n\frac{\partial}{\partial y^s}n_i  +  n_i \frac{\partial}{\partial y^s} \Delta u^n \\
&=(\tau^s n_i + \kappa^{\bot} s_i)  \Delta u^t  + t_i \frac{\partial}{\partial y^s} \Delta u^t \\
&- (\kappa^s s_i +\tau^s t_i )\Delta u^n  +  n_i \frac{\partial}{\partial y^s} \Delta u^n
\end{split}
\label{dev2}
\end{equation}

\subsection{Final expression of the regularised boundary element equations}
Using equations \eqref{dev0}, \eqref{dev1} and \eqref{dev2}, we obtain the final expression for the stress:

\begin{equation}
\begin{split}
\sigma_{ab}(\mathbf{x},t) &=c_{abcd}\int_{0}^{t}d\tau  \iint _{\Sigma}c_{ijpq}\frac{\partial}{\partial x_q} G_{cp} (n_d t_j- n_j t_d) \\ 
 &\left[  t_i \frac{\partial}{\partial y^t} \Delta u^t+n_i \frac{\partial}{\partial y^t}\Delta u^n\right]   d\Sigma  \\
&+c_{abcd}\int_{0}^{t}d\tau  \iint _{\Sigma}c_{ijpq}\frac{\partial}{\partial x_q} G_{cp}(n_d t_j- n_j t_d)  \\
& \left[  (\kappa^t  n_i+\kappa^{\parallel} s_i )\Delta u^t +  (\tau^t s_i - \kappa^t t_i )\Delta u^n  \right]   d\Sigma  \\
&+c_{abcd}\int_{0}^{t}d\tau  \iint _{\Sigma}c_{ijpq}\frac{\partial}{\partial x_q} G_{cp}(n_d s_j- n_j s_d) \\
& \left[ t_i \frac{\partial}{\partial y^s} \Delta u^t+n_i \frac{\partial}{\partial y^s}\Delta u^n   \right]   d\Sigma  \\
&+c_{abcd}\int_{0}^{t}d\tau \iint _{\Sigma}c_{ijpq}\frac{\partial}{\partial x_q} G_{cp} (n_d s_j- n_j s_d) \\
& \left[ (\tau^s n_i + \kappa^{\bot} s_i)  \Delta u^t  - (\kappa^s s_i +\tau^s t_i )\Delta u^n \right]   d\Sigma  \\
&-c_{abcd}\int_{0}^{t}d\tau \iint _{\Sigma} \frac{\partial^2}{\partial t^2}\Delta u_i(\mathbf{y},\tau)n_d \rho G_{ic}
\end{split}
\end{equation}

\section{Boundary-integral method:}
\subsection{Final expression for stresses in 2D static}
\label{stressB1}
\begin{equation}
\begin{split}
&\sigma_{ij}(\mathbf{x}) = \\
&\int_C K_{ij}^{\text{curv}}(\mathbf{x}-\mathbf{y}) \kappa^t\Delta u^t d L(\mathbf{y})+\int_C K_{ij}^{\text{grad}}(\mathbf{x}-\mathbf{y}) \frac{\partial}{\partial y^t}\Delta u^t d L(\mathbf{y})
\end{split}
\end{equation}

Where:
\begin{equation}
\begin{split}
K_{11}^{\text{grad}} &= \frac{\mu(1-p^2)}{\pi} \left[\left(\frac{\gamma_1}{r}-2\frac{\gamma_1\gamma_2^2}{r}\right)n_1(\mathbf{y})+\left(\frac{\gamma_2}{r}+2\frac{\gamma_1^2\gamma_2}{r}\right)n_2(\mathbf{y})\right] \\
K_{11}^{\text{curv}} &= \frac{\mu(1-p^2)}{\pi} \left[\left(\frac{\gamma_1}{r}-2\frac{\gamma_1\gamma_2^2}{r}\right)n_2(\mathbf{y})-\left(\frac{\gamma_2}{r}+2\frac{\gamma_1^2\gamma_2}{r}\right)n_1(\mathbf{y})\right] 
\end{split}
\end{equation}
\begin{equation}
\begin{split}
K_{22}^{\text{grad}} &=\frac{\mu(1-p^2)}{\pi}\ \left[\left(\frac{\gamma_1}{r}+2\frac{\gamma_2^2\gamma_1}{r}\right)n_1(\mathbf{y})+\left(\frac{\gamma_2}{r}-2\frac{\gamma_2\gamma_1^2}{r}\right)n_2(\mathbf{y})\right]  \\
K_{22}^{\text{curv}} &=\frac{\mu(1-p^2)}{\pi}\ \left[\left(\frac{\gamma_1}{r}+2\frac{\gamma_2^2\gamma_1}{r}\right)n_2(\mathbf{y})-\left(\frac{\gamma_2}{r}-2\frac{\gamma_2\gamma_1^2}{r}\right)n_1(\mathbf{y})\right] 
\end{split}
\end{equation}

\begin{equation}
\begin{split}
K_{12}^{\text{grad}} &=\frac{\mu(1-p^2)}{\pi} \left[(\gamma_1^2-\gamma_2^2)\frac{\gamma_2}{r}n_1(\mathbf{y})+(\gamma_2^2-\gamma_1^2)\frac{\gamma_1}{r}n_2(\mathbf{y})\right]  \\
K_{12}^{\text{curv}} &= \frac{\mu(1-p^2)}{\pi}\left[(\gamma_1^2-\gamma_2^2)\frac{\gamma_2}{r}n_2(\mathbf{y})-(\gamma_2^2-\gamma_1^2)\frac{\gamma_1}{r}n_1(\mathbf{y})\right] 
\end{split}
\end{equation}

These expressions match the ones by \cite{tada1997} for the effect of gradient slip.

\subsection{Tangential and normal traction}
We can get the tangential and normal traction by using:
\begin{equation}
\begin{split}
T_t(s) &= n_1(s)n_2(s)(\sigma_{11}-\sigma_{22})+(n_2^2(s)-n_1^2(s))\sigma_{12} \\
T_n(s) &= n_1^2(s)\sigma_{11}+n_2^2(s)\sigma_{22}+2n_1^2(s)n_2^2(s)\sigma_{12} 
\end{split}
\end{equation}

\begin{equation}
\begin{split}
T_t(s) &= \frac{\mu(1-p^2)}{\pi}\int d\mathbf{y} \frac{\partial}{\partial y}\Delta u_t(\mathbf{y},\tau) \\
&\times [4n_1(s)n_2(s)\gamma_1\gamma_2+(n_2^2(s)-n_1^2(s))(\gamma_2^2-\gamma_1^2)]\\
&\times \left( n_2(\mathbf{y})\frac{\gamma_1}{r} -n_1(\mathbf{y})\frac{\gamma_2}{r}\right) \\
&+ \frac{\mu(1-p^2)}{\pi}\int d\mathbf{y} \Delta u_t(\mathbf{y},\tau)  \kappa(\mathbf{y}) \\
&\times [4n_1(s)n_2(s)\gamma_1\gamma_2+(n_2^2(s)-n_1^2(s))(\gamma_2^2-\gamma_1^2)]\\
&\times \left( -n_1(\mathbf{y})\frac{\gamma_1}{r} -n_2(\mathbf{y})\frac{\gamma_2}{r}\right) 
\end{split}
\end{equation}
\begin{equation}
\begin{split}
T_n(s) &= \frac{\mu(1-p^2)}{\pi}\int d\mathbf{y} \frac{\partial}{\partial y}\Delta u_t(\mathbf{y},\tau) \\
& \times \{ n_1(\mathbf{y})\frac{\gamma_1}{r}+n_2(\mathbf{y})\frac{\gamma_2}{r}+ \\
&[2n_1(s)n_2(s)(\gamma_2^2-\gamma_1^2)-2(n_2^2(s)-n_1^2(s))\gamma_1\gamma_2]\\
&(n_2(\mathbf{y})\frac{\gamma_1}{r}-n_1(\mathbf{y})\frac{\gamma_2}{r})\} \\
&+ \frac{\mu(1-p^2)}{\pi}\int d\mathbf{y} \Delta u_t(\mathbf{y},\tau)  \kappa(\mathbf{y}) \\
& \times \{ n_2(\mathbf{y})\frac{\gamma_1}{r}-n_1(\mathbf{y})\frac{\gamma_2}{r}- \\
&[2n_1(s)n_2(s)(\gamma_2^2-\gamma_1^2)-2(n_2^2(s)-n_1^2(s))\gamma_1\gamma_2] \\
&(n_1(\mathbf{y})\frac{\gamma_1}{r}+n_2(\mathbf{y})\frac{\gamma_2}{r})\}
\end{split}
\end{equation}

\section{Numerical discretisation}
\begin{figure}
\centering
\includegraphics[width=75mm]{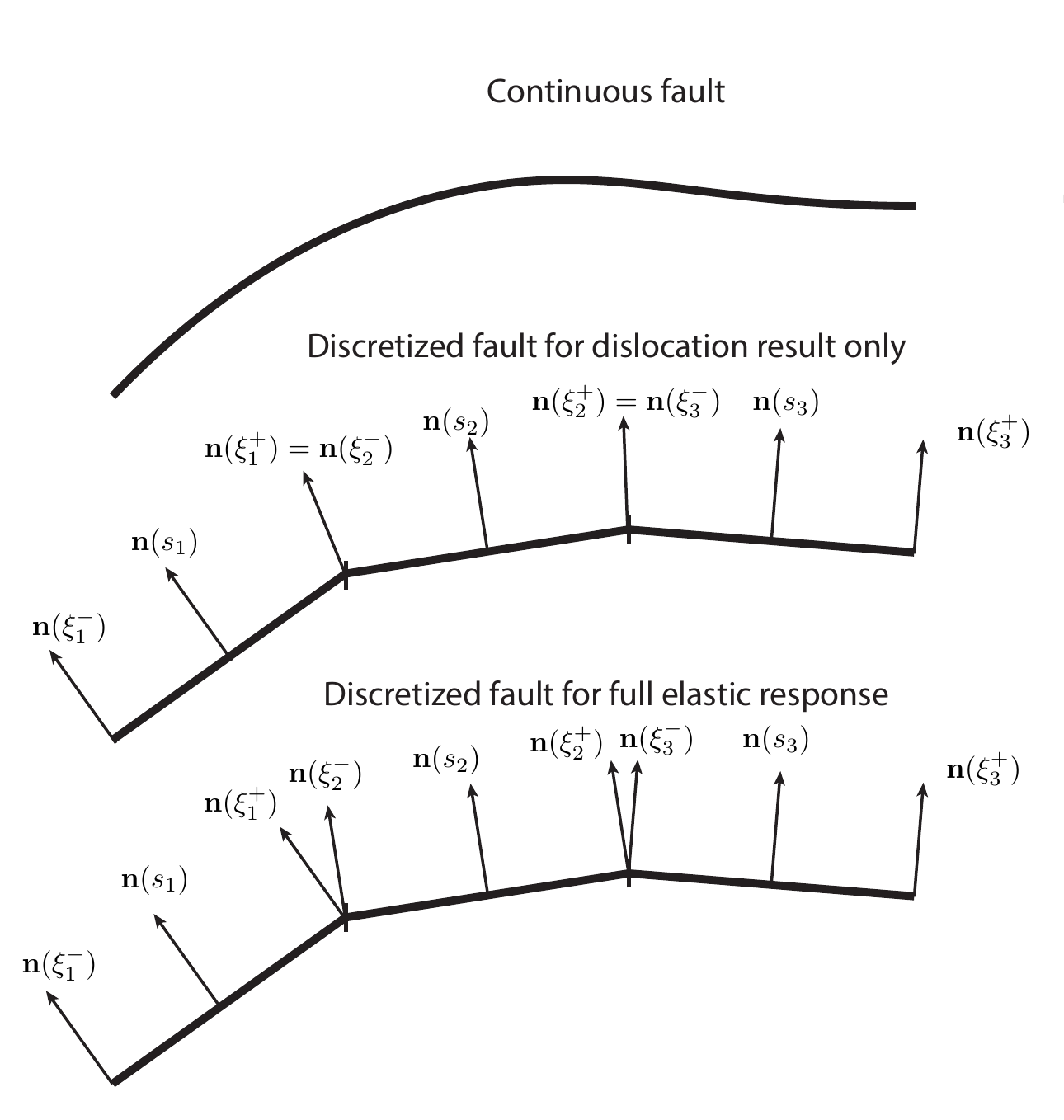}
\caption{Different discretisation of normal vector leads to different results.}
\label{BEM}
\end{figure}
To calculate the full elastic response, we considered piecewise constant slip over straight elements of size $ds$. The stress is then evaluated at the center of the element (\cite{rice1993,cochard1994}) (see Fig. \ref{BEM}). The point connecting two adjacent element is called a node.

\subsection{assumptions}
We assume piecewise constant for the slip as well as for the tangential vector.

\begin{equation}
\begin{split}
t^i(\xi)&=\sum_j  t_j^i (H(\xi-s_j+\Delta s/2)-H(\xi-s_j-\Delta s/2))\\
\Delta u^t(\xi)&=\sum_j\Delta u^t_j(H(\xi-s_j+\Delta s/2)-H(\xi-s_j-\Delta s/2))
\end{split}
\end{equation}

If we try to apply the previous expressions on a node point $s_k+\Delta s/2$, the answer is not straight forward because the tangent is not define between two adjacent straight element. However, if one assume that the Heaviside function follows $H(0)=0.5$, we can find a solution: 
\begin{equation}
\begin{split}
t^i(s_k+\Delta s/2)&=\sum_j  t_j^i (H(\xi-s_j+\Delta s/2)-H(\xi-s_j-\Delta s/2))\\
&= \frac{t_{k}^i+t_{k+1}^i}{2} 
\end{split}
\end{equation}
The same holds for the slip at a node point.
\subsection{Derivation}
Let's try to do the complete math for the whole problem:
\begin{equation}
\begin{split}
&\int K(s,\xi)\frac{\partial}{\partial x^t}(t_i \Delta u^t) = \int K(s,\xi)\left(\Delta u^t  \frac{\partial}{\partial x^t} t^i +t^i\frac{\partial}{\partial x^t}  \Delta u^t \right) \\
&\text{We include the discretisation} \\
&= \int K(s,\xi)\Delta u^t(\xi)  \\
&\frac{\partial}{\partial x^t}\sum_j  t_j^i (H(\xi-s_j+\Delta s/2)-H(\xi-s_j-\Delta s/2)) +\\
&  \int K(s,\xi) t^i(\xi) \\ 
&\frac{\partial}{\partial x^t}\sum_j  \Delta u^t_j(H(\xi-s_j+\Delta s/2)-H(\xi-s_j-\Delta s/2)) \\
&\text{We apply the derivative to the Heaviside function}\\
&=\int K(s,\xi)\Delta u^t(\xi)  \sum_j  t_j^i (\delta(\xi-s_j+\Delta s/2)-\delta(\xi-s_j-\Delta s/2)) +\\
& \int K(s,\xi)t^i(\xi)\sum_j  \Delta u_j^t (\delta(\xi-s_j+\Delta s/2)-\delta(\xi-s_j-\Delta s/2)) \\
\end{split}
\end{equation}
If we remove the integral and change the notation writting $s_j-\Delta s/2 = \xi_j$ and $s_j+\Delta s/2 = \xi_{j+1}$
\begin{equation}
\begin{split}
 &\sum_j  t_j^i (K(s,\xi_j)\Delta u^t(\xi_j)-K(s,\xi_{j+1})\Delta u^t(\xi_{j+1})) + \\
& \sum_j  \Delta u_j^t (K(s,\xi_j)t^i(\xi_j)-K(s,\xi_{j+1})t^i(\xi_{j+1}))
\end{split}
\end{equation}
Which is equivalent to
\begin{equation}
\begin{split}
 &\sum_j  t_j^i \left(K(s,\xi_j)\left[\frac{\Delta u^t_{j-1}+\Delta u^t_{j}}{2}\right]-K(s,\xi_{j+1})\left[\frac{\Delta u^t_{j}+\Delta u^t_{j+1}}{2}\right]\right) + \\
& \sum_j  \Delta u_j^t  \left(K(s,\xi_j)\left[\frac{t^i_{j-1}+t^i_{j}}{2}\right]-K(s,\xi_{j+1})\left[\frac{t^i_{j}+t^i_{j+1}}{2}\right]\right)\\
&= \sum_j  \Delta u_j^t  K(s,\xi_j)t^i_{j}- \Delta u_j^t  K(s,\xi_{j+1})t^i_{j}
\end{split}
\end{equation}
The other terms are canceling each other:
\begin{equation}
\begin{split}
t_j K(s,\xi_j) \Delta u_{j-1}\\
-t_j K(s,\xi_j) \Delta u_{j-1}
\end{split}
\end{equation}
We finally obtain:
\begin{equation}
\sigma_{ij}= \sum_j  \Delta u_j^t  K(s,\xi_j)t^i_{j}- \Delta u_j^t  K(s,\xi_{j+1})t^i_{j}
\end{equation}
Which is the expression we used to get the full elastic response. When showing only the curvature effect, we just discretised the fault so that the tangential vector at a node point is the mean of the tangential vector of two adjacent elements (see Fig. \ref{BEM}).

\bibliography{/Users/pierre/Dropbox/Public/CollectedPapers/MasterBibliography}
\bibliographystyle{apalike}

\end{document}